\documentclass[twocolumn]{aa}

\usepackage{graphicx}
\usepackage{txfonts}
\usepackage{natbib}

\begin{document}

   \title{3D numerical simulations of photodissociated and
photoionized disks}

   \author{M.J. Vasconcelos\inst{1}, A.H. Cerqueira\inst{1}
          \and
          A.C. Raga\inst{2}
          }

   \institute{Laborat\'orio de Astrof\'{\i}sica Te\'orica e Observacional, DCET \\
              Universidade Estadual de Santa Cruz, Rod. Ilh\'eus-Itabuna, km 16, Ilh\'eus, Bahia, Brasil \\
              \email{mjvasc@uesc.br, hoth@uesc.br}
         \and
             Instituto de Ciencias Nucleares, UNAM, Ap. Postal 70-543, CU, D.F. 04510, M\'exico \\
             \email{raga@nucleares.unam.mx}
             }


 
\abstract{}{In this work we study the influence of the UV radiation
field of a massive star on the evolution of a disklike mass of gas
and dust around a nearby star. This system has similarities with
the proplyds seen in Orion.} {We study disks with different
inclinations and distances from the source, performing fully 3D
numerical simulations. We use the YGUAZ\'U-A adaptative grid code
modified to account for EUV/FUV fluxes and non-spherical mass
distributions. We treat H and C photoionization in order to reproduce
the ionization fronts and photodissociation regions observed in
proplyds. We also incorporate a wind from the ionizing source, in
order to investigate the formation of the bow shock observed in
several proplyds. We examine density and H$\alpha$ maps, as well
as the mass loss rates in the photoevaporated winds.}{Our results
show that a photoevaporated wind propagates from the disk surface
and becomes ionized after an ionization front (IF) seen as a bright
peak in the H$\alpha$ maps. We follow the development of an
HI region inside the photoevaporated wind which corresponds to a
photodissociated region (PDR) for most of our models, except those
without a FUV flux. For disks that are at a distance from the source
$d \ge 0.1$ pc, the PDR is thick and the IF is detached from the
disk surface. In contrast, for disks that are closer to the source,
the PDR is thin and not resolved in our simulations. The IF then
coincides with the first grid points of the disk that are facing
the ionizing photon source. In both cases, the photoevaporated
wind shocks (after the IF) with the wind that comes from the ionizing
source, and this interaction region is bright in H$\alpha$.} {Our
3D models produce two emission features: a hemispherically shaped
structure (associated with the IF) and a detached bow shock where
both winds collide. A photodissociated region develops in all
of the models exposed to the FUV flux. More importantly, disks with
different inclinations with respect to the ionizating source have
relatively similar photodissociation regions. If the disk axis
is not aligned with the direction of the ionizing photon flux, the
IF displays moderate side-to-side asymmetries, in qualitative
agreement with images of proplyds, which also show such asymmetries.
The mass loss rates are $\sim 10^{-7}$ $M_{\odot}$ yr$^{-1}$ for
face-on disks, and $5\times 10^{-8}$ $M_{\odot}$ yr$^{-1}$ for
inclined disks at distances from 0.1 to 0.2 pc from the ionizing
photon sources.}

   \titlerunning{Photoevaporation of disks}
   \authorrunning{Vasconcelos, Cerqueira and Raga}

   \maketitle
%

\section{Introduction}\label{intro}

The ultraviolet radiation from a massive star in an environment of
low mass star formation produces several interesting phenomena,
among them the appearance of the so-called proplyds (an acronym
that stands for PROto-PLanetarY Disk; see O'Dell et al. 1993 and
O'Dell and Wen 1994).  Proplyds are young stellar objects (YSO)
with a cometary-shaped ionization front that contain a central, low
mass young star surrounded by an accretion disk, a photoevaporated
wind, and, in some cases, a monopolar or a bipolar outflow. The
proplyds in the Orion nebula (M42) were first reported as unresolved
optical emission features by \cite{laques..79} and then as radio
sources by \cite{church..87}. Using the Hubble Space Telescope
(HST), \cite{odell..94} and \cite{odell..96} reported the discovery
of tens of proplyds around the Trapezium cluster.

\begin{figure}
\centerline{\includegraphics[width=7cm]{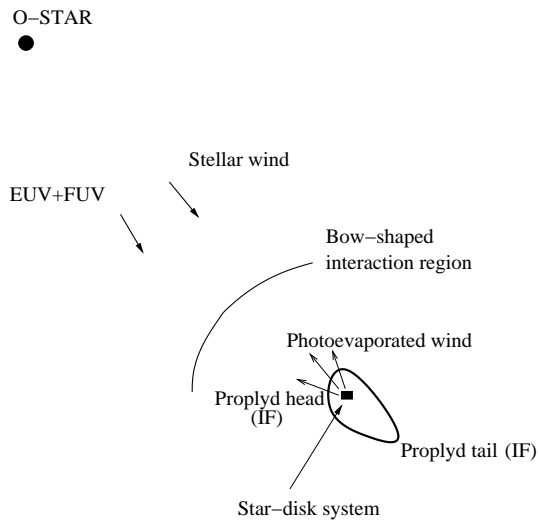}} 
\caption{A cartoon of a star-disk system being photoevaporated by
the UV radiation field from an O type star. The major features are
highlighted: the proplyd head and tail which traces the ionization
front from the direct ionization field coming from the O-star and
from the diffuse radiative field, respectively, and the interaction
region between the photoevaporated flow and the wind from the
O-star.}
\label{f1} 
\end{figure}

The proplyds present strong optical emission lines such as H$\alpha$,
[\ion{O}{III}] 5007 \AA, [\ion{O}{I}] 6300 \AA~ \citep{bally..00}
and also some ultraviolet \citep{henney..02} and IR lines
\citep{takami..02}. With HST high spatial resolution images and
spectra, it was possible to see that the emission comes from different
parts of the object. Proplyds present a cometary shaped feature,
very strong in H$\alpha$. Several of them show another bow shaped
feature, detached from the cometary structure itself, visible in
H$\alpha$, [\ion{O}{III}] 5007 \AA\ and in IR lines, lying further
from the low mass star than the ionization front (see Figure
\ref{f1}).  In some objects, the disks are seen in emission in
[\ion{O}{I}] 6300 \AA. It is also possible to see that the optical
lines are double or even triple peaked.  For example, in the central
parts of the proplyd 167-317 (LV2), H$\alpha$ has three components:
a low velocity component, peaking at a radial velocity of $\sim 30$
km s$^{-1}$, and two high velocity components, centered at $\sim
100$ km s$^{-1}$ and $\sim -75$ km s$^{-1}$ \citep{vasc..05}.

\cite{john..98} explained the observations with a model in which
the radiative field and the wind from $\theta^1$ Ori C interact
with a low mass YSO, generating a photoevaporated wind, an ionization
front (IF) and a bow shock. They also proposed the existence of a
photodissociated region (PDR), in which the photoevaporated wind
is neutral (the H is neutral). They separate their models in two
classes that depend on the amount of EUV and FUV flux arriving at
the proplyd's disk surface. The EUV model, valid for a YSO near to
or very far from $\theta^1$ Ori C ($d < 10^{17}$ cm or $d > 10^{18}$
cm), presents a very thin PDR. In this case, the EUV flux determines
the photoevaporated mass loss rate. On the other hand, for the FUV
model, the PDR is thick and can contain a shock front within it.
Here, the FUV flux determines the mass loss rate. \cite{storzer..99}
further improved the models by \cite{john..98} including results
from PDR codes.

Some axisymmetric numerical simulations were also done in order to
reproduce the characteristics of proplyds. \cite{garcia..01} carried
out axisymmetric simulations of the interaction of the stellar wind
from $\theta^1$ Ori C with the photoevaporated proplyd flow.  They
took available observational data to constrain stellar parameters
such as the wind density and velocity and the ionization front
parameters.  Their numerical calculations also assume a hemispherical
proplyd head and a cylindrical tail. Studying the interaction between
these two winds, they have successfully reproduced the arc emission
for the proplyds near $\theta^1$ Ori C.

\cite{rich..00} performed 2D, axisymmetric hydrodynamical simulations
of photoevaporating disks including both ionizing (h$\nu \geq 13.6$
eV) and dissociating (6 eV $<$ h$\nu <$ 13.6 eV) radiation. In their
models, disk structures are formed through collapse simulations of
1 and 2 $M_{\odot}$ molecular clumps \citep{yorke..99}, which are
then exposed to the radiation field by switching on the external
UV radiation field in the calculation. They studied the effects of
distance from the UV photon source on emission line maps and the
effect of the presence of a spherical wind from the proplyd star,
which promotes the appearance of collimated, bipolar microjets.
Instead of calculating the dissociation of the H$_2$ molecule, they
followed the ionization of C. They took into account photoelectric
heating and cooling by fine-structure lines such as [\ion{C}{II}]
158 $\mu$m, [\ion{O}{I}] 63 $\mu$m and [\ion{O}{I}] 145 $\mu$m. An
important point in their work is the inclusion of the diffuse
radiation field, which is responsible for the tails of the proplyds
(see Figure \ref{f1}).

In this work we present the first fully three-dimensional numerical
simulations of disks exposed to FUV and EUV radiation fields.  As
discussed in \cite{john..98}, the diffuse ionizing and dissociating
radiation field is important for determining the shape of the
ionization front behind the (directly) illuminated disk surface.
Our present simulations do not include the diffuse field, resulting
in the tails that do not have the correct morphology (see Cerqueira
et al. 2006b). We concentrate on obtaining a description of the
head of the proplyd flow, and study the effect of different
orientations of the disks with respect to the impinging UV photon
field (and the stellar wind).

The present paper is organized as follows.  In \S \ref{simulations},
we describe the code and the simulations. In \S \ref{results}, we
present our results. Finally, in \S \ref{conclusions}, we draw our
main conclusions.

\begin{figure}
\centerline{\includegraphics[width=8cm]{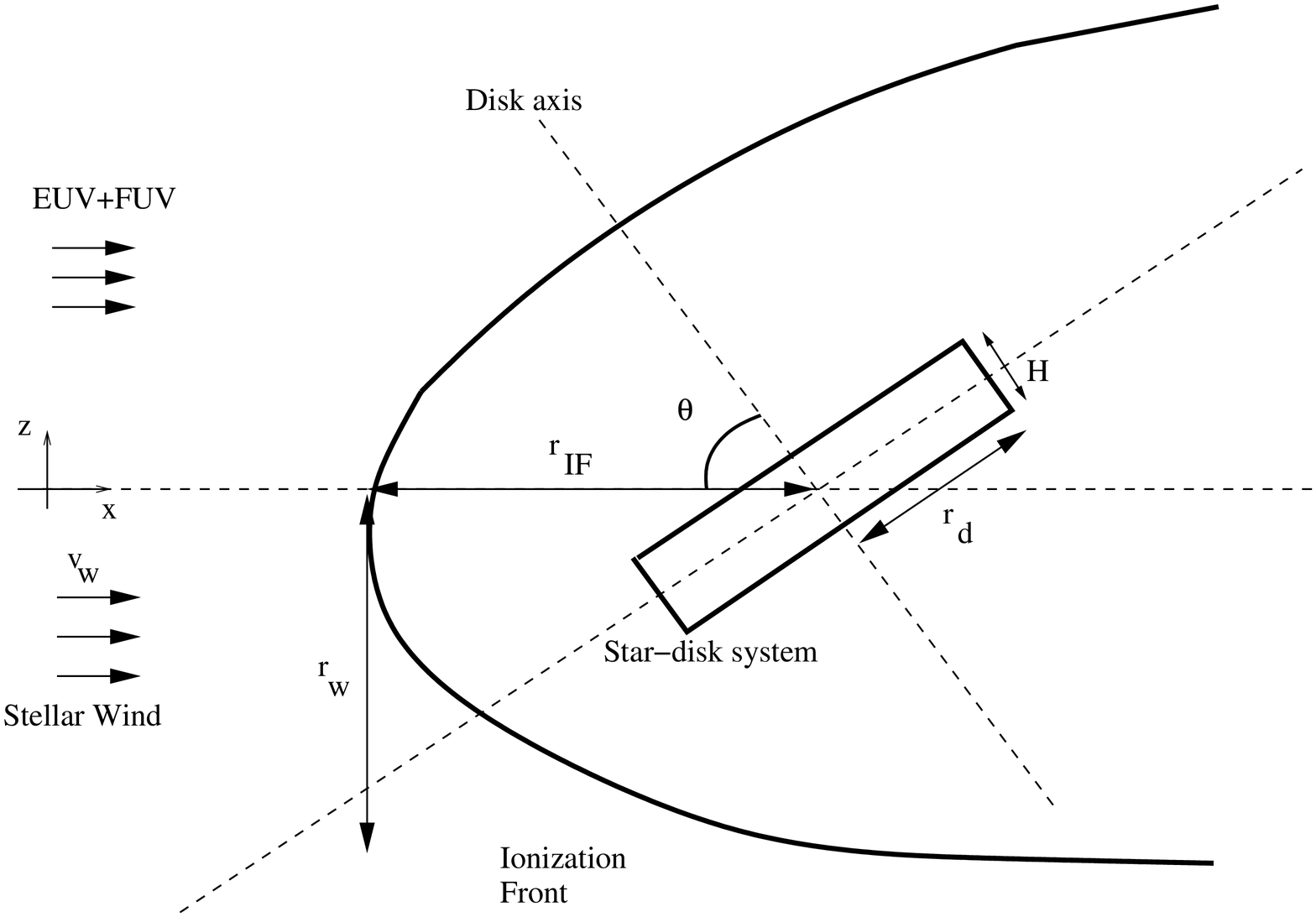}} 
\caption{Schematic view of the initial setup of the simulations.
The cartoon shows also the distance from the disk to the ionization
front, $r_{IF}$, and the definition of the width of the ionized
structure, $r_w$.  We follow the notation of \cite{john..98} (see
Figure 7 in that paper).}
\label{f2}
\end{figure}

\begin{table}
\begin{minipage}[t]{\columnwidth}
\caption{Computed models}
\label{tab1}
\centering
\renewcommand{\footnoterule}{}  
\begin{tabular}{cccccc}
\hline \hline
Model &  S$_\mathrm{FUV}$ & d & $\theta$\footnote{Inclination of
the disk, in degrees, with respect to the $z$-axis.} &
v$_w$\footnote{Velocity of the wind from the O star.} & T$_2$\footnote{The
temperature associated with the PDR.} \\
 & ($\times 10^{49}$ s$^{-1}$) & (pc)
& & (km s$^{-1}$) & (K) \\
\hline
M1a  & 0    & 0.1  & $0^{\circ}$ & 0   & 1\,000   \\
M1b  & 0    & 0.1  & $0^{\circ}$ & 100 & 1\,000   \\
M2a & 1.78 & 0.2  & $0^{\circ}$ & 100 & 1\,000   \\
M2b & 1.78 & 0.1  & $0^{\circ}$ & 100 & 1\,000   \\
M2c & 1.78 & 0.02 & $0^{\circ}$ & 100 & 1\,000   \\
M2d & 1.78 & 0.1  & $0^{\circ}$ & 100 & 3\,000   \\
M3a & 1.78 & 0.2  & $45^{\circ}$ & 100 & 1\,000   \\
M3b & 1.78 & 0.1  & $45^{\circ}$ & 100 & 1\,000   \\
M4a & 1.78 & 0.2  & $75^{\circ}$ & 100 & 1\,000   \\
M4b & 1.78 & 0.1  & $75^{\circ}$ & 100 & 1\,000   \\
\hline
\end{tabular}
\end{minipage}
\end{table}

\section{The numerical method, the numerical setup, and the simulated
models} \label{simulations}

\begin{figure*}
\centering
\includegraphics[width=19cm]{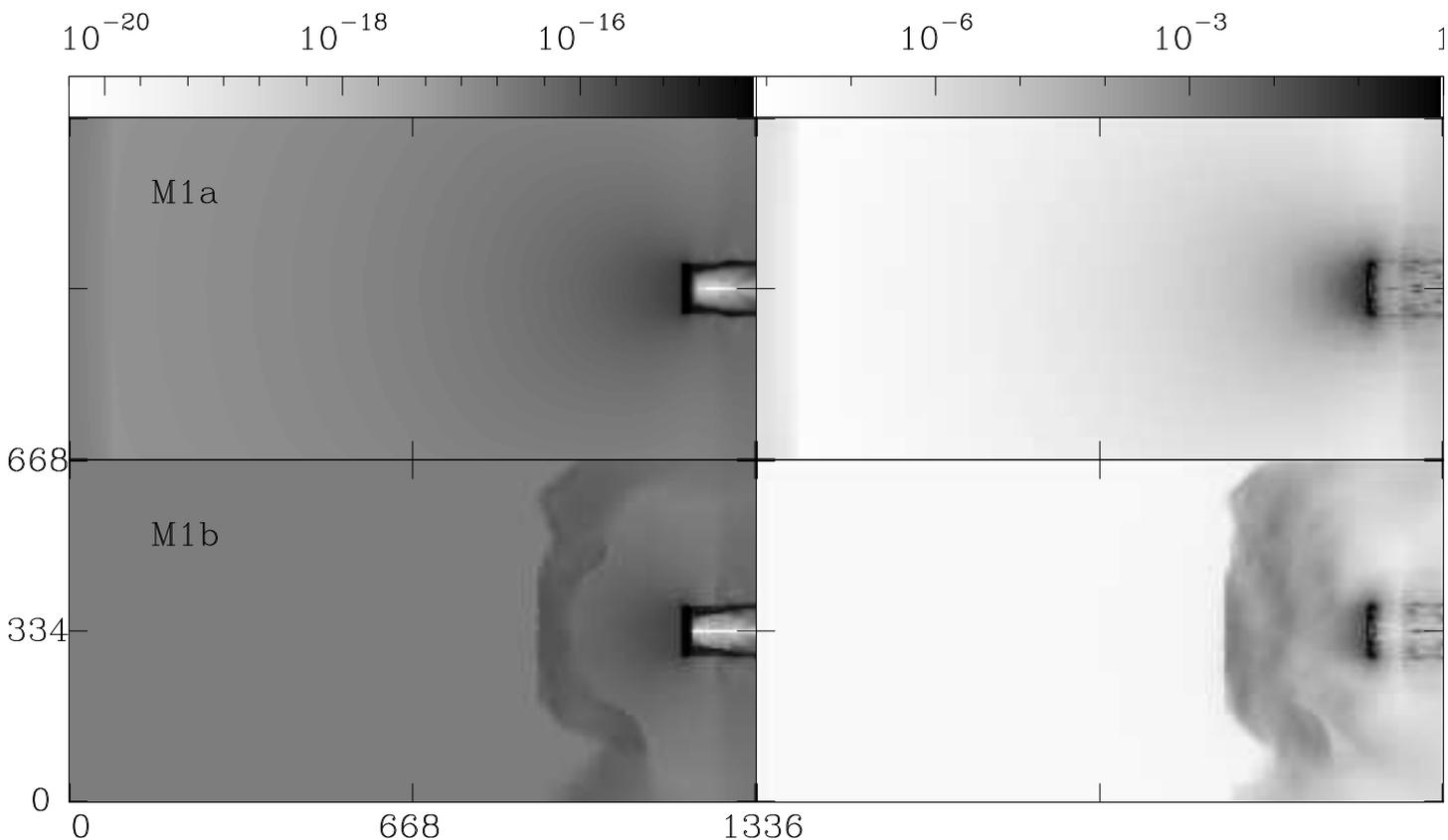} 
\caption{Density $xz$-midplane stratifications (left) and H$\alpha$
maps (right) obtained from models M1a (top) and M1b (bottom) which
has only an EUV radiation field for $t=2\,000$ yr. The scales (given
by the top-bars) are in g cm$^{-3}$ for the density (left), and
normalized to one for the H$\alpha$ maps (right). The coordinate
axes are in AU.}
\label{f3} 
\end{figure*}

The 3D numerical simulations have been carried out with the YGUAZ\'U-A
adaptive grid code \citep{raga..00,raga..02} using a 5-level binary
adaptive grid. The YGUAZ\'U-A code integrates the gasdynamic equations
employing the flux vector splitting scheme of \cite{vanleer82}
together with a system of rate equations for atomic/ionic species.
In our simulations, we consider 4 species: \ion{H}{I}, \ion{H}{II},
\ion{C}{I} and \ion{C}{II}. This code has been extensively employed
for simulating different astrophysical flows such as jets
\citep{masciadri..02, dri..06a}, interacting winds \citep{ricardo..04},
photoevaporating clumps \citep{dri..06b} and supernova remnants
\citep{pablo..01a,pablo..04}. It was also tested with laser generated
plasma laboratory experiments \citep{sobral..00,raga..01,pablo..01b}.

In this work, instead of solving an energy equation, we prescribe
a temperature law, given by

\begin{equation} \label{temp}
T = (T_1 - T_2)\cdot x_\mathrm{HII} + T_2\cdot x_\mathrm{CII} +
T_3\cdot (1 - x_\mathrm{CII}),
\end{equation}

\noindent where $T_1 = 10\,000$ K, is the characteristic temperature
of an \ion{H}{II} region, $T_2 = 1\,000$ K, the typical PDR
temperature, $T_3 = 10$ K, the temperature of the molecular gas,
$x_\mathrm{H II}$ is the hydrogen ionization fraction and $x_\mathrm{C
II}$ is the carbon ionization fraction ($x_\mathrm{C II}=1$ when
all the $\mathrm{C I}$ is $\mathrm{C II}$). This prescription is
justified if the thermal equilibrium time scale is much smaller
than the dynamical time scale \citep{lefloch..94}, which is the
case here, for both the ionized and the molecular gas.  It is clear
from equation (\ref{temp}), that we have possible temperatures
ranging from 10 K to $10^4$ K.  In order to study the dependence
of the PDR geometry with temperature, we also compute models in
which we set $T_2 = 3\,000$ K.

Following \cite{rich..00} we do not treat the dissociation of H$_2$,
but consider that the dissociation front and the carbon ionization
front coincide. We then solve rate equations for hydrogen and carbon
taking into account photoionization, collisional ionization, radiative
and dielectronic recombination.

We model the O star as a point source located outside the computational
domain. This source emits EUV photons (with h$\nu > 13.6$ eV) at a
rate $S_{\rm EUV}=7.2\times10^{48}$ s$^{-1}$ and FUV photons (with
6 eV $<$ h$\nu < 13.6$ eV) at a rate $S_{\rm FUV}=1.78\times10^{49}$
s$^{-1}$ (or zero, in order to study the effects of the EUV radiation
field only; see \S \ref{results} below).

For the transfer of FUV photons, we consider the optical depth due
to the ionization edge of the C~I photoionization cross section,
and to the absorption of photons by dust:

\begin{equation}\label{tau-fuv}
\tau_{FUV}={1.22\times 10^{-17}{\rm cm^2}}N_{CI}+ {1.15\times
10^{-21}{\rm cm^2}}N_{HI}\,,
\label{tauf}
\end{equation}

\noindent where $N_{CI}$ is the C~I column density integrated along
the path of the ray. The second term (representing the FUV dust
absorption) is calculated assuming that the optical extinction is
given by $A_V=N_{HI}/{2\times 10^{21}{\rm cm^{-2}}}$ (appropriate
for a standard grain distribution). We then multiply this number
by 2.4 to obtain the FUV extinction. This is consistent with
empirically derived optical/UV extinction curves (see, e.~g.,
Cardelli et al. 1988). We are also implicitly assuming that there
is no dust in the region where H is fully ionized.

\begin{figure}
\centering
\includegraphics[width=9cm]{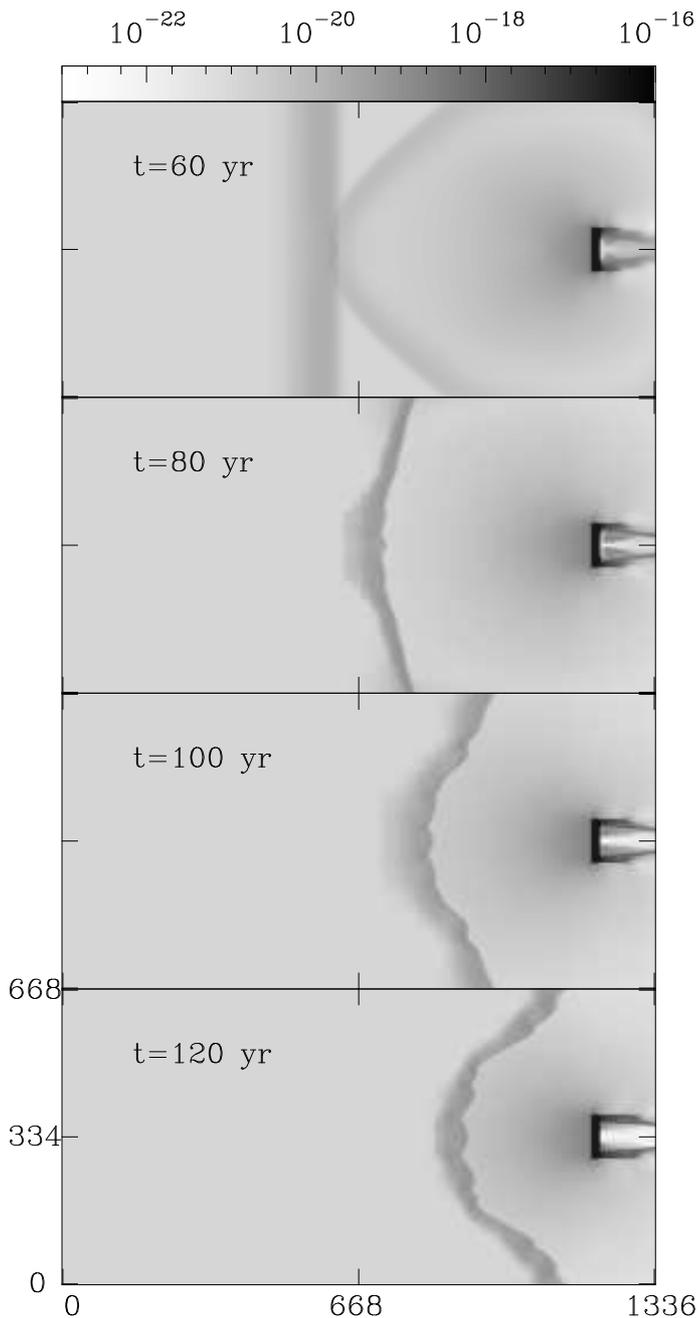}
\caption{Density $xz$-midplane stratifications for model M1b, which
has only an EUV radiation field, from $t=80$ yr (top) to $t=120$
yr (bottom), showing the interaction between the stellar wind with
the photoevaporated flow. The density scale (given by the top-bar)
is in g cm$^{-3}$. The coordinate axes are in AU.}
\label{f4} 
\end{figure}

Our simulations include a flat, neutral disk, with a constant
thickness $H = $ 15.6 AU and an outer radius of 50 AU. The disk is
positioned at distances of 0.02, 0.1 and 0.2 pc from the ionizing
photon source. With this particular choice of parameters, we can
in principle ensure that the photoevaporated flow of the nearest
disk would be EUV-dominated, while for the more distant one,
FUV-dominated.  Using previous analytical calculations of \cite{john..98}
and the disk radius given above, we can roughly estimate the minimum
distance from the ionizing source in order to have an FUV-dominated
flow as $d_{min} \sim 0.1$ pc. The disk radius that would support
an FUV-dominated flow is, for $d = 0.1$ pc, $r_d \approx 30$ AU,
and $r_d \approx 120$ AU for $d=0.2$ pc (see equations 15 and 16
in Johnstone {\it et al.} 1998). We choose to simulate disks at
distances from the ionizing photon source that are i) at a distance
that is suitable for an EUV-dominated flow ($d=0.02$ pc), ii) close
to the limiting cases for which we should expect an FUV dominated
flow to occur ($d=0.1$pc, $r_d=50$AU), and iii) FUV-dominated flow
($d=0.2$pc).  In this last case, the disk radius $r_d = 50$ AU is
found to obey the $r_d \lesssim 122$ AU condition, and the system
would develop an FUV-dominated flow. The only caveat is that, as
pointed out by \cite{john..98}, there are several sources of
uncertainties in evaluating these (analytical) numbers.

The density of the disk is assumed to be constant in both the
vertical and radial directions. We start the simulations assuming
pressure equilibrium between the PDR region and the disk itself:

\begin{equation}\label{density}
n_d=n_0(T_2/T_3),
\end{equation}

\noindent where $T_2$ and $T_3$ are the PDR and disk temperatures
(see Equation \ref{temp}) and $n_0$ is the number density at the
base of the photoevaporated flow, which can be estimated as:

\begin{equation}\label{base}
n_0 \simeq N_D/r_d .
\end{equation}

\noindent for a FUV dominated flow. Here, $N_D$ is the column density
of the PDR attained for $\tau_{FUV} \sim 1-3$, that is of the order
of 10$^{21}$ cm$^{-2}$ \citep{john..98}. Using $r_d=50$ AU, we
obtain $n_0 \sim 10^6$ cm$^{-3}$.  We therefore set $n_d=10^7$
cm$^{-3}$. For an EUV dominated flow, the density is found by
equating the EUV flux to the recombinations in the ionized flow:

\begin{equation}\label{base2}
\frac{S_{EUV}}{4\pi d^2}=3\alpha_r n_0^2 r_d\,,
\end{equation}

\noindent where $\alpha_r=2.6\times10^{-13}$ cm$^3$ s$^{-1}$ is the
recombination coefficient for Hydrogen at 10$^4$ K, $d$ is the
distance from the disk to the ionizing photon source, and
$S_{EUV}=7.2\times10^{48}$ s$^{-1}$ and $r_d=50$ AU in all of the
models presented in this paper. We should note that the factor of
3 on the right hand side of equation (\ref{base2}) is a result of
assuming a spherical divergence for the photoevaporated wind.

We then calculate the densities $n_0$ for different distances and
find $n_0=(0.98-5.0)\times10^5$ cm$^{-3}$ for $d=(0.1-0.02)$ pc.
For all cases, we treat the disk as an infinite reservoir, since
at each time step we reset its initial conditions, as described
above.

We assume the presence of a 0.2$M_{\odot}$ star at the disk center
and consider the gravitational force that it exerts on the flow (by
including the appropriate source terms in the momentum and energy
equations). A Keplerian rotation velocity law (consistent with the
0.2~M$_\odot$ central star) is imposed on the disk material.  The
gravitational force is important as it inhibits the formation of a
photoevaporated wind in the central regions of the disk.  We expect
a photoevaporated flow to start beyond the gravitational radius,
which can be estimated as $r_g = GM_{\star}/a^2$, where $a$ is the
local sound speed. For FUV dominated flows, $a \simeq 3$ km s$^{-1}$,
so that $r_g = 2.9\times10^{14}$ cm, or $\approx 20$ AU.  The
inhibition of the photoevaporated wind in the central regions of
the disk can be seen in our FUV dominated flow simulations (which
have $r_g\approx 20$~AU), but not in our EUV dominated flow simulations
(which have $r_g\approx 2$ AU, smaller than our numerical resolution).
The size of the remnant disk, which is known to be typically smaller
than the gravitational radius, will not be investigated in this
work, since we are treating the disk as an infinite reservoir
\citep{john..98}.

For some models, we add a plane-parallel wind (flowing parallel to
the ionizing/dissociating external radiation field, with velocity
$v_w = 100$ km s$^{-1}$ and with number density $n_w = 500$ cm$^{-3}$),
which represents the wind from the O-type star. The wind velocity
of $\theta^1$ Ori C is estimated to be of the order of 500 to
$1\,650$ km/s \citep{bally..98}, or even larger (up to about $3\,600$
km s$^{-1}$, Walborn \& Nichols 1994), which is by far larger than
the adopted velocity here. On the other hand, the ram pressure
estimated for both the stellar wind and the photoevaporated flow
for proplyds in the Trapezium region is $\sim 3 \times10^{-8}$ dyn
cm$^{-2}$ to 5$\times10^{-7}$ dyn cm$^{-2}$ for the proplyds closer
to $\theta^1$ Ori C (at distances $\simeq 0.01 pc$; see, e.g.,
Garcia-Arredondo {\it et al.} 2001). Therefore, our particular
choice of parameters for the stellar wind give us a ram pressure
comparable to that estimated for a group of proplyds in the Trapezium
cluster. In particular, with the adopted parameters, we have $P_{sw}
= 8.3\times10^{-8}$ dyn cm$^{-2}$.

\begin{figure*}
\centering
\includegraphics[width=15cm]{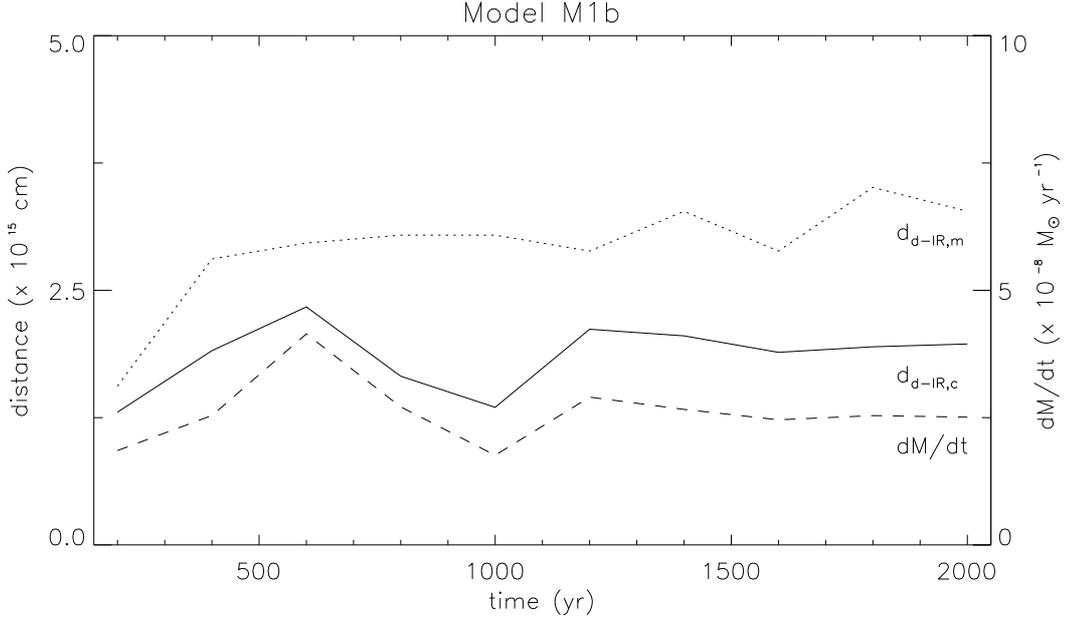}
\caption{Distances from the disk to the interaction region as a
function of time are depicted for model M1b, which has only an EUV
radiation field: $d_{{\rm{d-IR,c}}}$ (full line) are the calculated
distances, where stellar and the photoevaporated wind ram pressures
match; $d_{{\rm{d-IR,m}}}$ (dotted-line) are the distances as
measured from the simulations (along the $x$-axis, with
$y=z=4.9\times10^{15}$ cm). The distances are in units of $10^{15}$
cm (left-axis). Also shown is the disk mass loss $\dot{M}$
(dashed-line), in units of $10^{-8}$ M$_{\odot}$ yr$^{-1}$ (right
axis).  The $\dot{M}$ has been calculated by computing the quantity
$\rho\vec{A}\cdot \vec{v}$ on the faces of a cube centered at the
disk center, with $a=3r_d$.}
\label{f5} 
\end{figure*}

The initial setup of the simulations is shown in Fig. \ref{f2}. The
flat, dense structure with which we model the circumstellar disk
is placed on the $yz$-plane, at $x=$ 1$\,$203 AU.  The disk axis,
which we define here as the axis perpendicular to the disk plane,
and that contains the disk center at ($x$,$y$,$z$)= (1203,334,334)
AU, coincides with the $x$-axis for an inclination angle
$\theta=0^{\circ}$. For larger inclination angles, the disk axis
rotates clockwise, with respect to the $x$-axis, as depicted in
Figure \ref{f2}.  The remaining volume of the computational domain
is filled with an ionized environment of uniform density $n_a =
500$ cm$^{-3}$. All models except M1a have an initial wind of
v$_\mathrm{w}$ = 100 km s$^{-1}$ and density n$_\mathrm{w} = 500$
cm$^{-3}$, entering the computational domain through the $x=0$
boundary. The ionizing/dissociating photon source is placed along
the $x$-axis, outside the computational domain to the left.

In Table \ref{tab1} we present all the relevant parameters used in
our models: the dissociating photon rate, $S_{FUV}$ (in units of
$10^{49}$ s$^{-1}$; second column), the distance from the disk to
the external star (in pc; third column), the inclination angle
$\theta$ between the disk and the $x$ axis (fourth column), the
velocity of the wind from the external star (in km s$^{-1}$; fifth
column) and the temperature of the PDR region (in K; sixth column).
All the models have an ionizing photon flux $S_{EUV}=7.2\times
10^{48}$~s$^{-1}$ (similar to the values used by Richling \& Yorke
2000) and $r_d = 50$ AU.

The simulations have been computed in a 5-level, binary adaptive
grid with a maximum resolution (along the three axes) of 5.2 AU for
all models. The computational domain has a size
$(x_{max},y_{max},z_{max})=$(1$\,$337,668,668) AU.

\section{Results and Discussion} \label{results}

In Fig. \ref{f3}, we show the $xz$-midplane density stratifications
(left) and H$\alpha$ maps (right) for model M1a  (top) and model
M1b (bottom) at $t=2\,000$ years. The models present only an EUV
radiation field from the ionizing source (at a 0.1 pc distance from
the disk). The difference between the two models is the presence
of a stellar wind in model M1b, as described in Table \ref{tab1}.
The disk axis has no inclination with respect to the direction of
the impinging photons (i. e., $\theta=0^{\circ}$ see Figure \ref{f2}
and also Table \ref{tab1}).

The H$\alpha$ emission in model M1a, is restricted to a region near
the disk surface (and facing the ionizing source), and the edge of
the shadow on the non-illuminated side of the disk.  For model M1b,
we have H$\alpha$ emission close to the disk (with morphology and
intensity similar to those seen in model M1a), and also fainter
emission regions distributed along the interaction region between
the stellar wind and the photoevaporated flow. It is known that
some proplyds in Orion show faint H$\alpha$ arcs facing $\theta^1$
Ori C \citep{bally..98}, which correspond to the stellar
wind/photoevaporated flow interaction emission found in our models.

\begin{figure*}
\centering
\includegraphics[width=18cm]{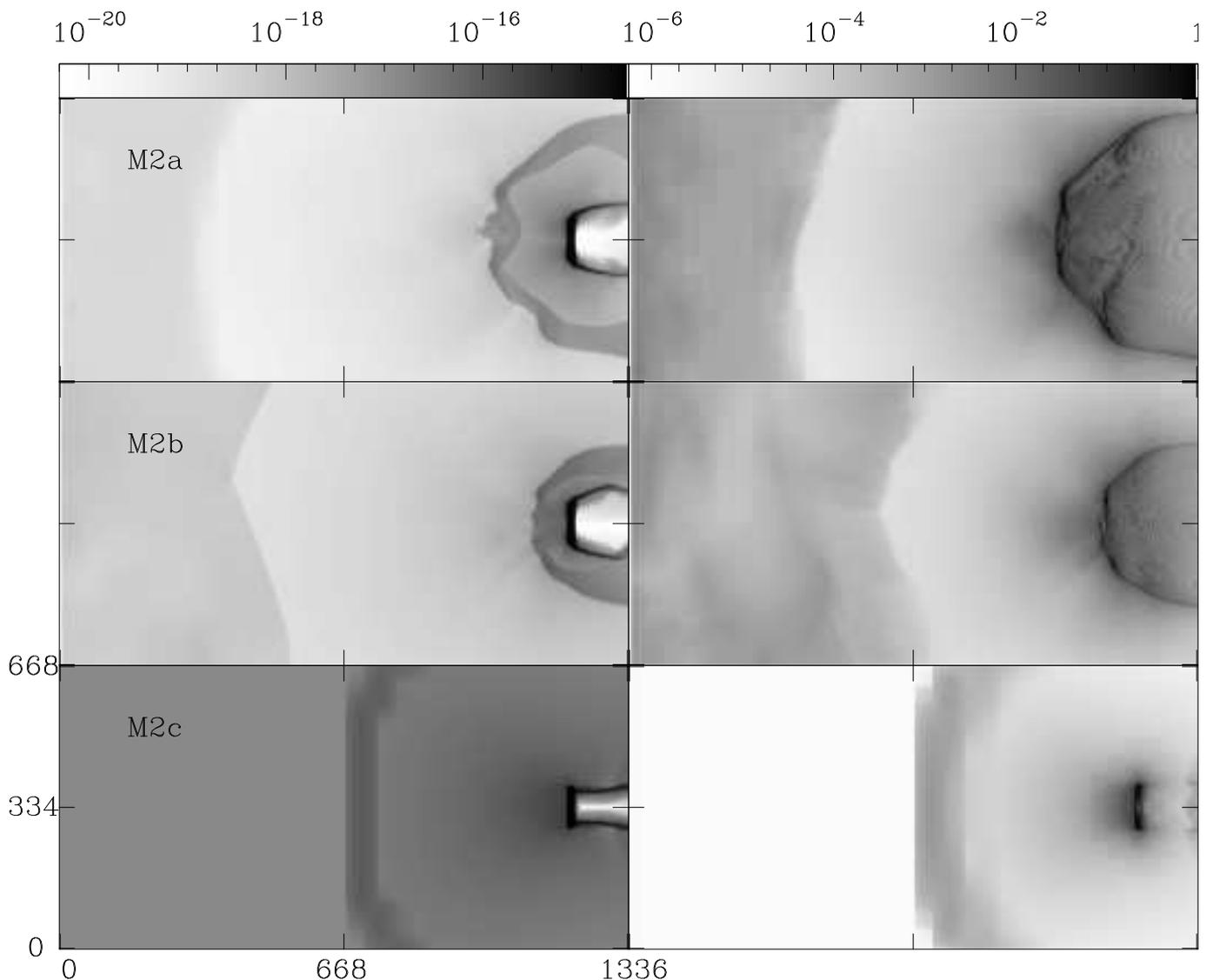}
\caption{Density $xz$-midplane stratifications (left) and H$\alpha$
maps (right) for models M2a (top), M2b (middle) and M2c (bottom)
at $t=2\,000$ yr. The scales (given by the top bars) are in g
cm$^{-3}$ for the density (left), and normalized to one for the
H$\alpha$ maps (right). The coordinate axes are in AU.}
\label{f6} 
\end{figure*}

In Figure \ref{f4} we show $xz$-midplane density maps for model M1b
at $t=$ 60, 80, 100 and 120 yr, in order to illustrate the initial
evolution of such an interaction. At $t=60$ yr, the stellar wind
and the photoevaporated flow start to interact at $x \simeq$ 668
AU. At later times, the interaction region moves towards the disk,
and stabilizes around the position of the point of ram pressure
balance between the stellar wind and the photoevaporated flow.  We
can also see this effect in Figure \ref{f5}, where we show the
photoevaporated wind mass loss rate  ($\dot{M}$; dashed line), the
distance from the disk to the interaction region i) as measured
directly from the $xz$-midplane density maps ($d_{\rm{d-IR,m}}$;
dotted-line) and ii) as calculated ($d_{\rm{d-IR,c}}$; full-line)
by balancing the ram pressures from the stellar wind ($P_{sw}$) and
from the photoevaporated flow ($P_{pef}$).  The expression for
$d_{\rm{d-IR,c}}$, that takes into account the logarithmic outward
acceleration expected for the photoevaporated wind is given by (see
Henney et al. 1996):

\begin{equation}\label{ramp}
d_{{\rm{d-IR,c}}} \simeq R_0\cdot \bigg[1+ 4{\rm ln}\bigg(
\frac{R_0}{r_d} \bigg) \bigg]^{1/4},
\end{equation}

\noindent where $r_d$ is the disk radius, and $R_0$:

\begin{equation}\label{ramp2}
R_0 = \sqrt{\frac{\dot{M}a_{{\rm II}}}{4\pi P_{sw}}}.
\end{equation}

\noindent In this equation, $a_{{\rm II}}$ is the sound speed of
the photoionized flow, and $P_{sw}$ is the ram pressure of the
stellar wind, which is $P_{sw} = 8.3\times10^{-8}$ dyn cm$^{-2}$
for our chosen parameters (see \S \ref{simulations}).  In Figure
\ref{f5}, we see that both the measured and the calculated distances
are comparable.

In Fig. \ref{f6}, we show the $xz$-midplane density stratifications
(left) and H$\alpha$ maps (right) for models M2a (top), M2b (middle)
and M2c (bottom), at $t=2\,000$ years.  The models have both EUV
and FUV radiation fields (from the stellar source). The distance
from the disk to the ionizing source is 0.2 pc, 0.1 pc and 0.02 pc
for models M2a, M2b and M2c, respectively; see Table \ref{tab1}.
The density of the disk is determined from equations (\ref{density})
and (\ref{base}) for the FUV dominated models M2a and M2b, and
equations (\ref{density}) and (\ref{base2}) for the EUV dominated
model M2c. The disk axis is aligned with the $x$-direction
($\theta=0^{\circ}$ see Figure \ref{f2} and also Table \ref{tab1}),
and the stellar wind is present in the three models.

\begin{figure*}
\centering
\includegraphics[width=16cm]{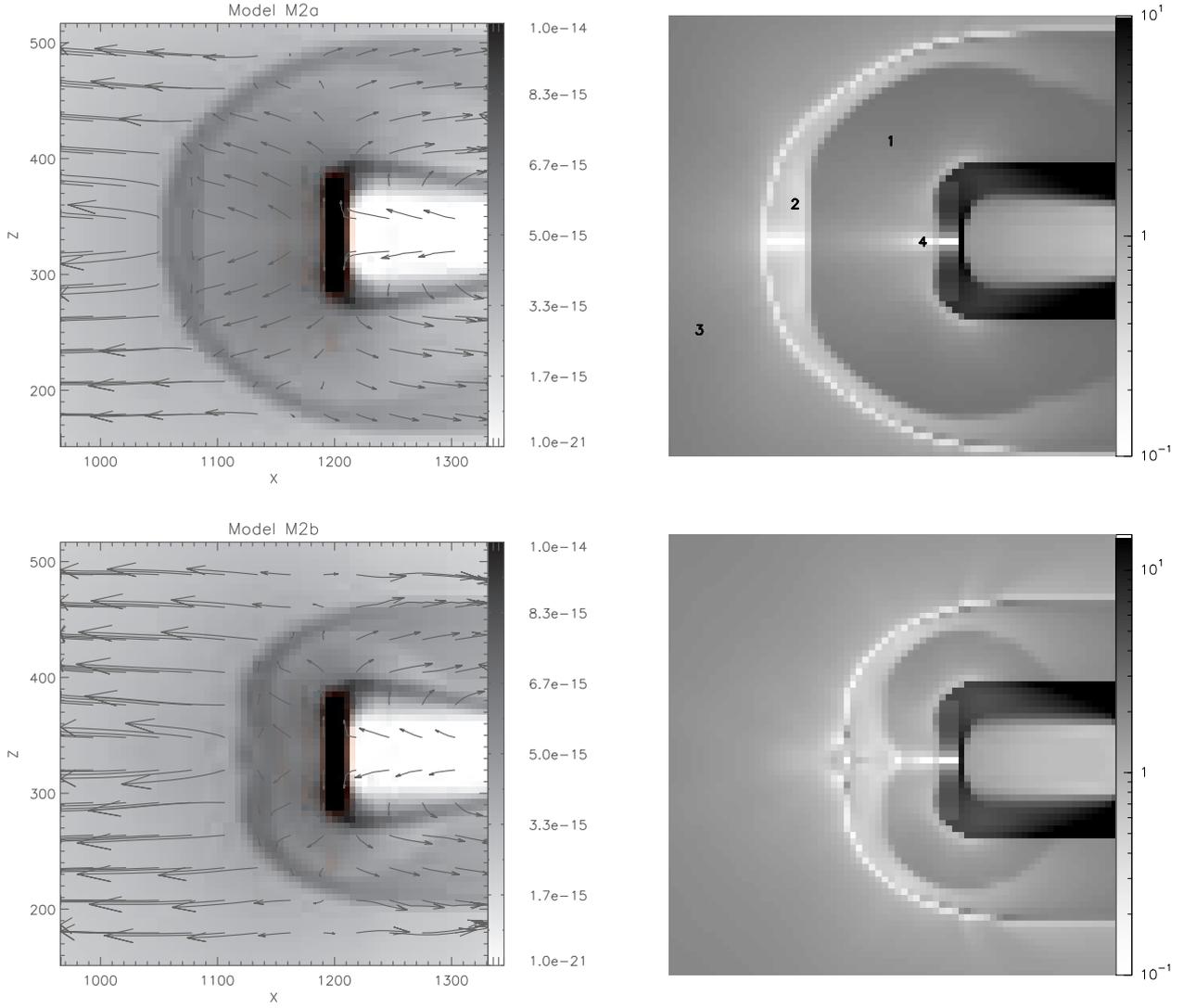} 
\caption{Density $xz$-midplane stratifications superimposed on
velocity field vectors (left panels) and an xz-map of the Mach
number (right panels), $M=v/a$ ($a$ is the local sound speed) for
models M2a (top) and M2b (bottom), at $t = 200$ yr. The snapshots
are a closer view of the star-disk system being photoevaporated.
The distance (in AU) to the origin of the coordinate system is
depicted for the $x$ and $z$ axes in the left panels. The right-hand
side bars on each panel indicate the density scale (in g cm$^{-3}$;
left panels) and the Mach number (right panels). The labels in the
top-rigth panel indicates the region where we have a supersonic
outflow (1 and 3), a subsonic outflow (2), and a subsonic inflow
(4). We also note that region (1) is neutral, at a temperature
$T_2=10^3$ K, whereas (3) is ionized, at a temperature $T_1 = 10^4$
K.}
\label{f7} 
\end{figure*}

We see that a high density region shrouds the disk in models M2a
and M2b. This region is related to the photodissociation region
expected from the analytical models of \cite{john..98} and
\cite{storzer..99}. It has a temperature $T_2 \simeq 1\,000$ K (see
equation \ref{temp} and Table \ref{tab1}). This region is composed
of \ion{C}{II} and \ion{H}{I}.  We also note that the environment
(the left part of the computational domain) and the photoevaporated
wind outside the PDR have a temperature of about $10^4$ K, indicating
that they are photoionized. The photodissociation region separates
the disk from the ionized photoevaporated wind.  In model M2c (in
which the disk is at 0.02 pc from the ionizing photon source) the
photodissociation region is limited to a thin shell in contact with
the surface of the disk.

The ionization front radius (as defined in Figure \ref{f2}) is found
to grow from $r_{IF}=2.4r_d = 120$AU at $t=200$ yr to $r_{IF}=3.8r_d
= 190$AU at $t=2\,000$ yr for model M2a, while for model M2b,
$r_{IF}=1.6r_d=80$ AU (constant throughout the simulation). In
models M2a and M2b, the outer radius of the photodissociation region
($r_w$, defined in Figure 2) grows with time.  This lateral expansion
occurs at a velocity of $\approx 0.1$~km~s$^{-1}$, at $\sim 1/10$
of the $\sim 1$~km~s$^{-1}$ outwards flow velocity of the gas within
the photodissociation region.  This indicates that the ionization
front in these models  is pressure confined by the base of the
photoevaporated flow. The ratio between the cross section radius
$r_w$ and the axial standoff distance $r_{IF}$ (see Figure 2) has
values $r_w/r_{IF} = 1.4$ and $1.9$ for models M2a and M2b (at
$t=2\,000$ yr), respectively.  Interestingly, \cite{john..98} predict
$r_w=1.3 r_{IF}$ from their simplified, analytic model.  Models M2a
and M2b have detached, almost hemispherical ionization fronts. On
the other hand, in model M2c the ionization front lies very close
to the surface of the disk (with just one pixel of separation between
them). It is interesting to note that observations of proplyd
170-337 \citep{john..98}, that is situated at $~0.1$ pc from
$\theta^1$ Ori, show that its IF radius is approximately 80 AU.
From the analytical calculations of \cite{john..98} for FUV dominated
flows, r$_\mathrm{IF14} \simeq 0.1 (\epsilon^2/f_rS_{49})d_{17}^2$,
where r$_\mathrm{IF14}$ is the radius of the IF in units of $10^{14}$
cm, $d_{17}$ is the distance from the source in units of $10^{17}$
cm, $S_{49}$ is the ionizing photon rate in units of 10$^{49}$
s$^{-1}$, $f_r$ is the fraction of EUV photons absorbed by recombination
in the ionized region of the flow, and $\epsilon$ is a dimensionless
parameter of order unity \citep[see][]{john..98}. Taking
$(\epsilon^2/f_rS_{49})=16$ \citep{john..98}, and d$_{17} = 3.08$,
we obtain  r$_\mathrm{IF} \simeq 100$ AU, which is consistent with
the values obtained from models M2a and M2b.

In Figure \ref{f7} we show the velocity field in a region close to
the accretion disk superimposed on density (left panels) and Mach
number maps (right panels), $M=v/a$, where $a$ is the local sound
speed and $v$ is the local velocity of the gas, for models M2a (top
panels) and M2b (bottom panels), at $t=200$ yr. From Figure \ref{f7},
we see that the photoevaporated flow has four different dynamical
regimes, indicated in the top-right panel of Figure \ref{f7} by the
labels 1 to 4, which refer to: 1) a supersonic neutral outflow, 2)
a subsonic, neutral outflow, both of them inside the PDR region and
separated by a shock; 3) a supersonic, ionized outflow outside the
PDR region and 4) a subsonic, neutral inflow, localized around the
disk center (at $r \lesssim r_g$).

\begin{figure*}
\centering
\includegraphics[width=18cm]{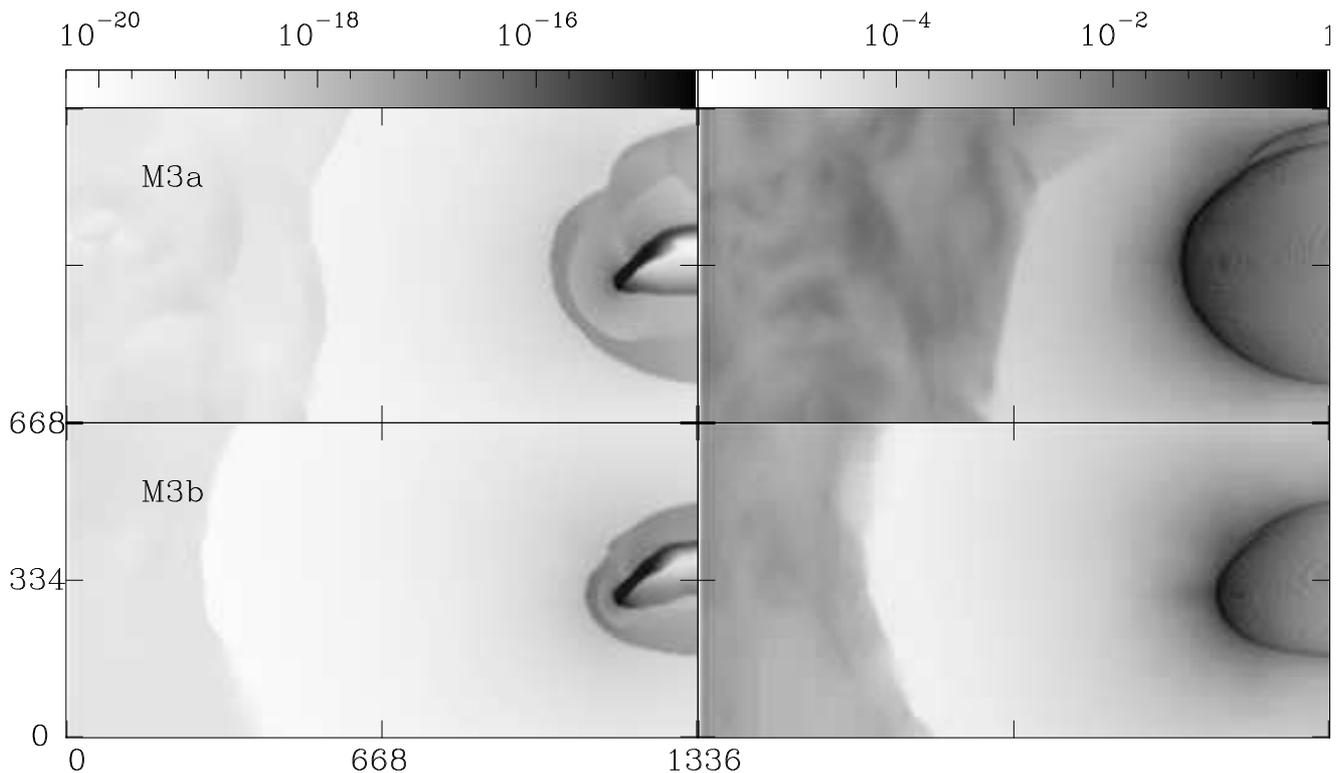}
\caption{Density $xz$-midplane stratifications (left) and H$\alpha$
maps (right) obtained from models M3a (top) and M3b (bottom) for
$t=2\,000$ yr. The scales (given by the top bars) are in g cm$^{-3}$
for the density (left), and normalized to one for the H$\alpha$
maps (right). The coordinate axes are in AU. The distance between
the ionized photon source and the disk is $d=0.2$ pc and $d=0.1$
pc for models M3a and M3b, respectively. In both cases, $\theta=45^{\circ}$
(see Figure 2).}
\label{f8} 
\end{figure*}

As we have stated before in \S \ref{simulations}, the disk radius
adopted in our models is larger than the gravitational radius.  For
instance, for the parameters of the M2 models (M2a, M2b and M2c),
we have $r_g \simeq 20$ AU.  We can see from the velocity field
maps that between disk radial distances $r=0$ (the disk axis) and
$r\simeq \pm 3\Delta z$ (where $\Delta z= 5.2$ AU is the spacing
of the highest resolution grid, see \S \ref{simulations}), we do
not have an outflow.  For $r \gtrsim 20$ AU, the outflow starts
supersonically from the disk surface, with Mach numbers $\approx
1.1$ (region 1 in Figure \ref{f7}), and reach an internal shock
(before the IF; see Figure \ref{f7}) with Mach numbers $\approx
1.8$ (note that the PDR has a sound velocity $a_I = 3.2$ km s$^{-1}$).
Between this internal shock and the outer boundary of the PDR, the
outflow is subsonic, with Mach numbers $\approx$ 0.4-0.6 (region 2
in Figure \ref{f7}).  Outside the PDR and before reaching the
interaction region between the two winds, we have a supersonic
outflow with Mach numbers $\approx$ 1.2-1.6 (with a sound velocity
$a_{II} = 14.2$ km s$^{-1}$).

The structure obtained from models M2a and M2b is similar to the
scenario proposed by \cite{john..98} for the FUV dominated flow.
From this we conclude that both the M2a and M2b models produce
photoevaporated flows in the FUV dominated regime.

\begin{figure*}
\centering
\includegraphics[width=18cm]{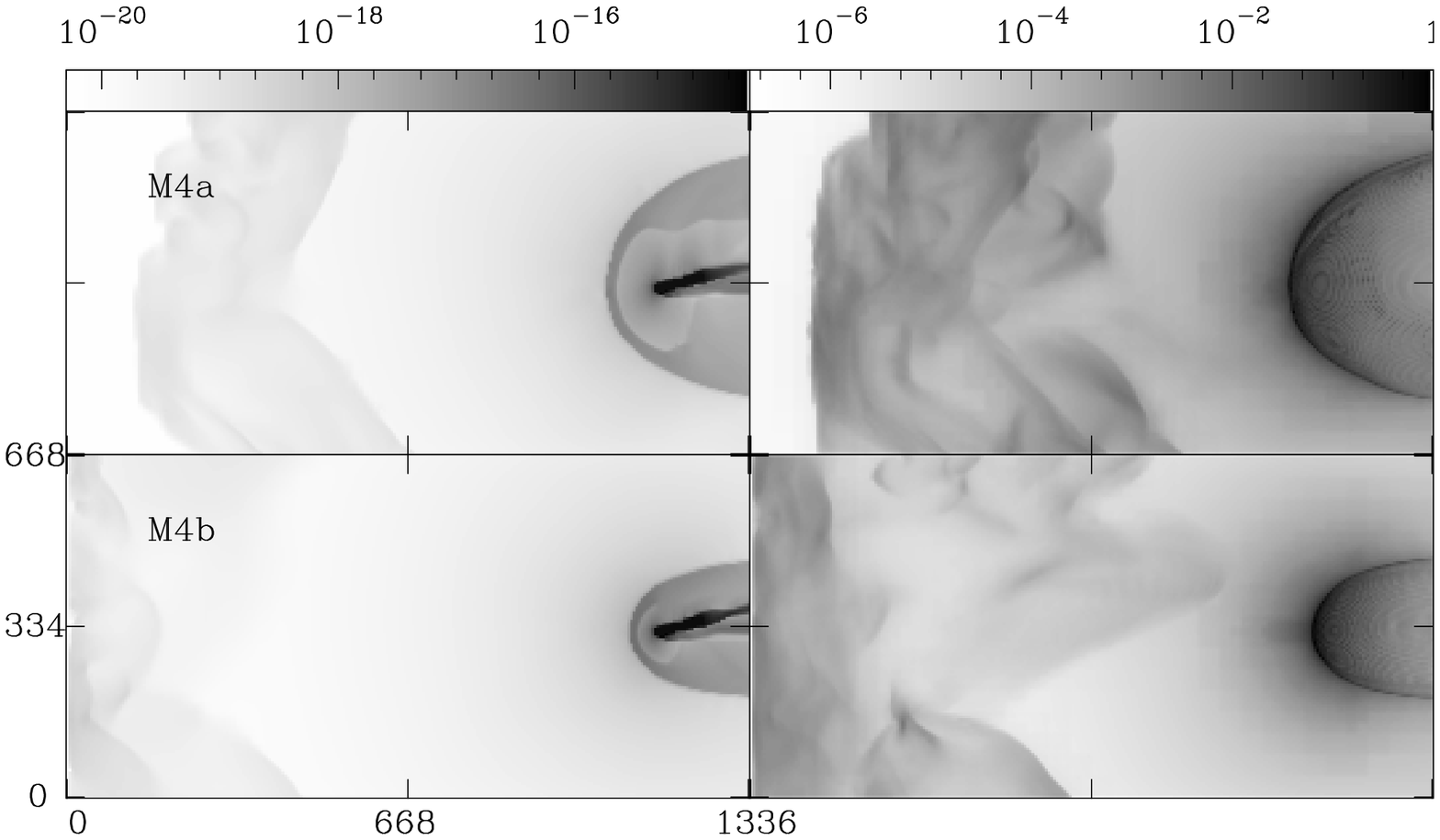}
\caption{Density $xz$-midplane stratifications (left) and H$\alpha$
maps (right) obtained from models M4a (top) and M4b (bottom) for
$t=2\,000$ yr. The scales (given by the top bars) are in g cm$^{-3}$
for the density (left), and normalized to one for the H$\alpha$
maps (right).  The coordinate axes are in AU.}
\label{f9} 
\end{figure*}

In Figure \ref{f8} we show the $xz$-midplane density stratifications
(left) and H$\alpha$ maps (right) for models M3a (top) and M3b
(bottom panels) at $t=2\,000$ yr. The models have an EUV+FUV radiation
field from the ionizing source, that is located at 0.2 pc (M3a) and
0.1 pc (M3b) distance from the disk.  In both cases, the disks have
a 45$^{\circ}$ of inclination with respect to the direction of the
impinging photons ($\theta=45^{\circ}$ see Figure \ref{f2} and Table
\ref{tab1}).  In Figure \ref{f9} we present the same results, but
for models M4a and M4b (with $\theta=75^{\circ}$). As can be seen
in Table \ref{tab1}, models M2a, M3a and M4a differ only by the
values of the disk inclination. The same is true for the set of
models M2b, M3b and M4b.

\begin{figure*}
\centering
\includegraphics[width=16cm]{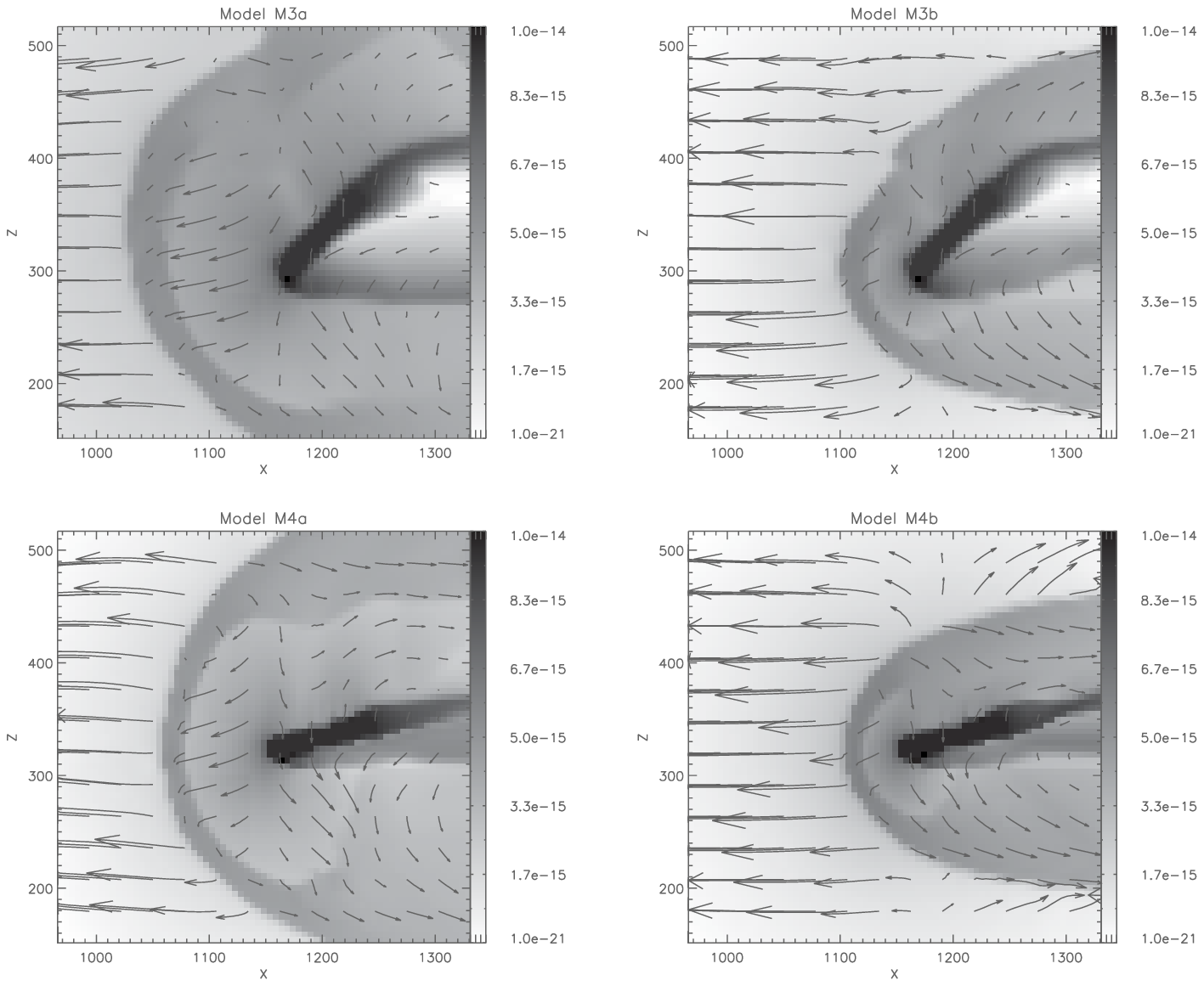}
\caption{Density $xz$-midplane stratifications superimposed on
velocity field vectors for models M3a (top-left), M3b (top-right),
M4a (bottom-left) and M4b (bottom-right), at $t = 2\,000$ yr.  The
distance to the origin of the coordinate system, in AU, is depicted
for the $x$ and $z$ axes.  The right-hand side bars indicate the
density scale (in g cm$^{-3}$).}
\label{f10}
\end{figure*}

In the inclined disk models (models M3a, M3b, M4a and M4b, see
Figures 8-9) the internal structure of the PDR region is the same
as in the M2a and M2b models (with $\theta=0^{\circ}$, see Figure
\ref{f7} and Table 1). There is a neutral supersonic outflow (with
$M\approx 2$ in all models) near the illuminated surface of the
disk. For inner disk radial distances (from the disk axis), there
is a gas inflow. In Figure \ref{f10} we show velocity vectors
superimposed on density maps for models M3a (top-left), M3b
(top-right), M4a (bottom-left) and M4b (bottom-right). As in the
M2a and M2b models, we again have an internal shock that separates
the supersonic from the subsonic neutral flows, inside the PDR.
Outside this internal shock we have the outer edge of the PDR region,
that separates the subsonic, neutral flow from the external, ionized
supersonic flow. These structures are characteristic of an FUV
dominated flow, as discussed previously.

The H$\alpha$ emission in the region close to the ionization front
of models M3 and M4 (Figures \ref{f8} and \ref{f9}) has a general
morphological resemblance to the H$\alpha$ morphology of the
``face-on'' models (M2, see Figure \ref{f6}). However, the H$\alpha$
maps of the inclined disk models show clear side-to-side asymmetries,
which are more pronounced for the models with $\theta=45^\circ$
(M3a, b, see Table 1 and Figure \ref{f8}) than for the ones with
$\theta=75^\circ$ (M4a, b, see Figure \ref{f9}).

From the $t=2\,000$~yr time-frames of models M3 and M4 (see Figures
8 and 9), we have calculated the axial stand-off distance ($r_{IF}$)
and width of the cross section presented by the ionization front
($r_w$, see Figure 2). We obtain the following results:

\begin{itemize}

\item M3a: $r_{IF}=3.6r_d$ = 180 AU; $r_w=4.4r_d$ = 220 AU,
\item M3b: $r_{IF}=2.0r_d$ = 100 AU; $r_w=2.4r_d$ = 120 AU,
\item M4a: $r_{IF}=2.8r_d$ = 140 AU; $r_w=3.7r_d$ = 185 AU,
\item M4b: $r_{IF}=1.9r_d$ = 95 AU; $r_w=2.1r_d$ = 105 AU,

\end{itemize}

\noindent with $r_w/r_{IF} \lesssim 1.3$ for all models.  These
results illustrate the reduction in the size of the photodissociation
region that is obtained for increasing values of the disk inclination
angle $\theta$.

\begin{figure*}
\centering
\includegraphics[width=18cm]{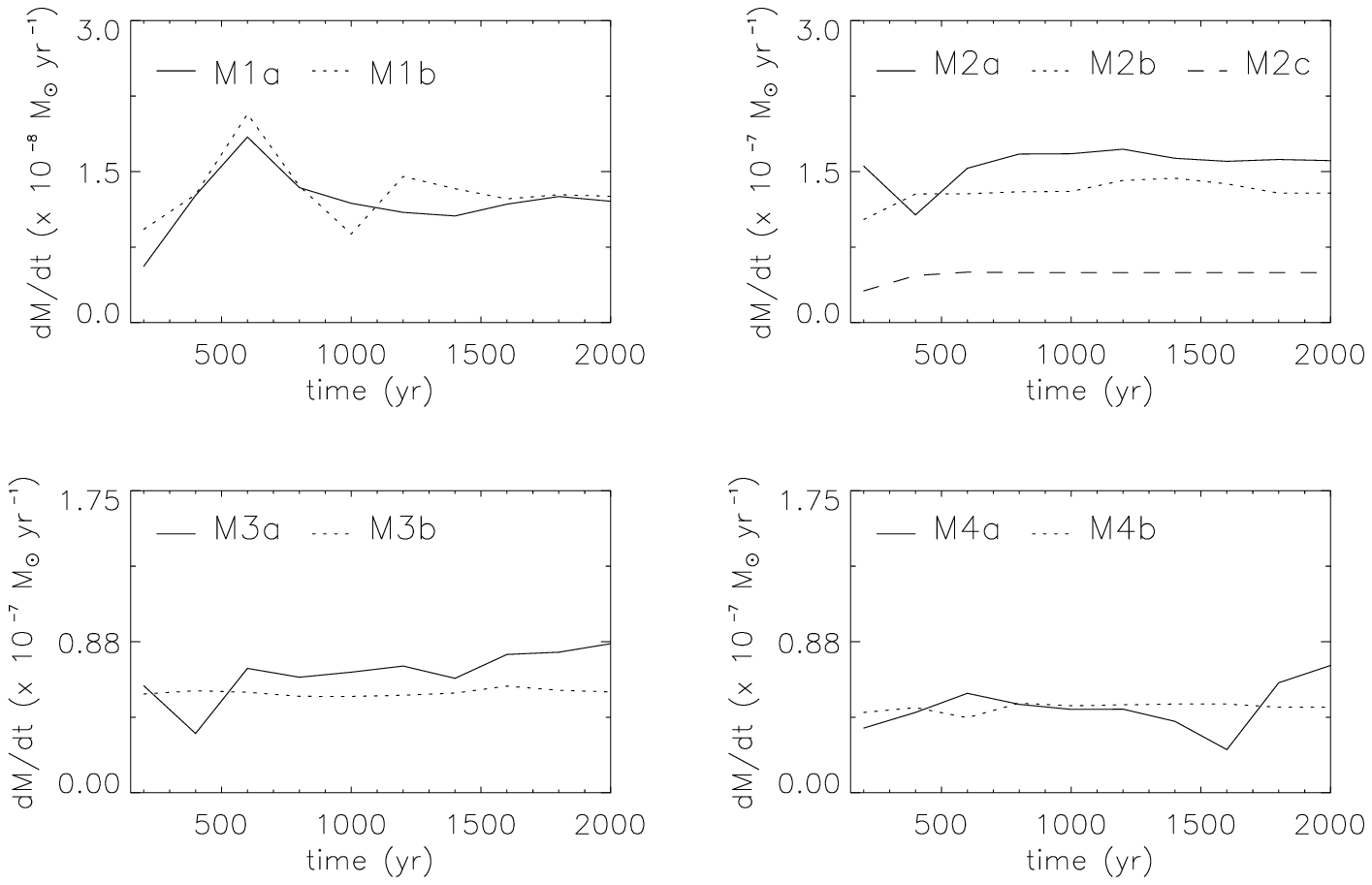}
\caption{The mass loss rate as a function of time for models M1a
and M1b (top-left panel), M2a, M2b and M2c (top-right panel), M3a
and M3b (bottom-left panel) and M4a and M4b (bottom-right panel).
The caption on the top of each diagram indicates the model that a
given curve type represents. The mass loss rate is in units of
$10^{-7}$ M$_{\odot}$ yr$^{-1}$ for models M2, M3 and M4 and in
units of $10^{-8}$ M$_{\odot}$ yr$^{-1}$ for models M1.}
\label{f11}
\end{figure*}

In Figure \ref{f11} we present the mass loss rate as a function of
time, for all the simulated models except model M2d. The mass loss
rate for models M1a and M1b (without an FUV radiation field; see
Table \ref{tab1}) shows a small time variability (as mentioned
above), but is almost independent of the presence of the stellar
wind.  Also, the mass loss rate has approximately the same value
even if the distance from the disk to the ionizing photon source
changes by a factor of two, even for the are FUV dominated models,
as is the case for those with a given inclination angle (with respect
to the ionizing photon source) as M2a and M2b ($\theta=0^{\circ}$),
M3a and M3b ($\theta=45^{\circ}$), and M4a and M4b ($\theta=75^{\circ}$).
This is due to our simplified model, in which the PDR temperature
is assumed to be position independent and with the same value for
all models. Looking at a series of models with increasing values
of the inclination angle $\theta$ (e. g., models M2a, M3a and M4a,
which otherwise have identical paramenters, see Table 1), we see
that the mass loss rate as a monotonically decreasing function of
$\theta$. This result is to be expected, given the lower effective
cross sections (to the impinging UEV/FUV radiation field) presented
by the disk for increasing inclination angles. In particular, if
we take the mean value of $\dot{M}$ for models M2a and M2b (with
$\theta=0$), for models M3a and M3b (with $\theta=45^{\circ}$) and
for models M4a and M4b (with $\theta=75^{\circ}$), we find $\dot{M}_{M2}
= 1.4\times10^{-7}$ M$_{\odot}$ yr$^{-1}$, $\dot{M}_{M3} =
6.2\times10^{-8}$ M$_{\odot}$ yr$^{-1}$$\dot{M}_{M4} = 4.8\times10^{-8}$
M$_{\odot}$ yr$^{-1}$, which implies $\dot{M}_{M3} \simeq 0.4
\dot{M}_{M2}$ and $\dot{M}_{M4} \simeq 0.3 \dot{M}_{M2}$.

\begin{figure*}
\centering
\includegraphics[width=18cm]{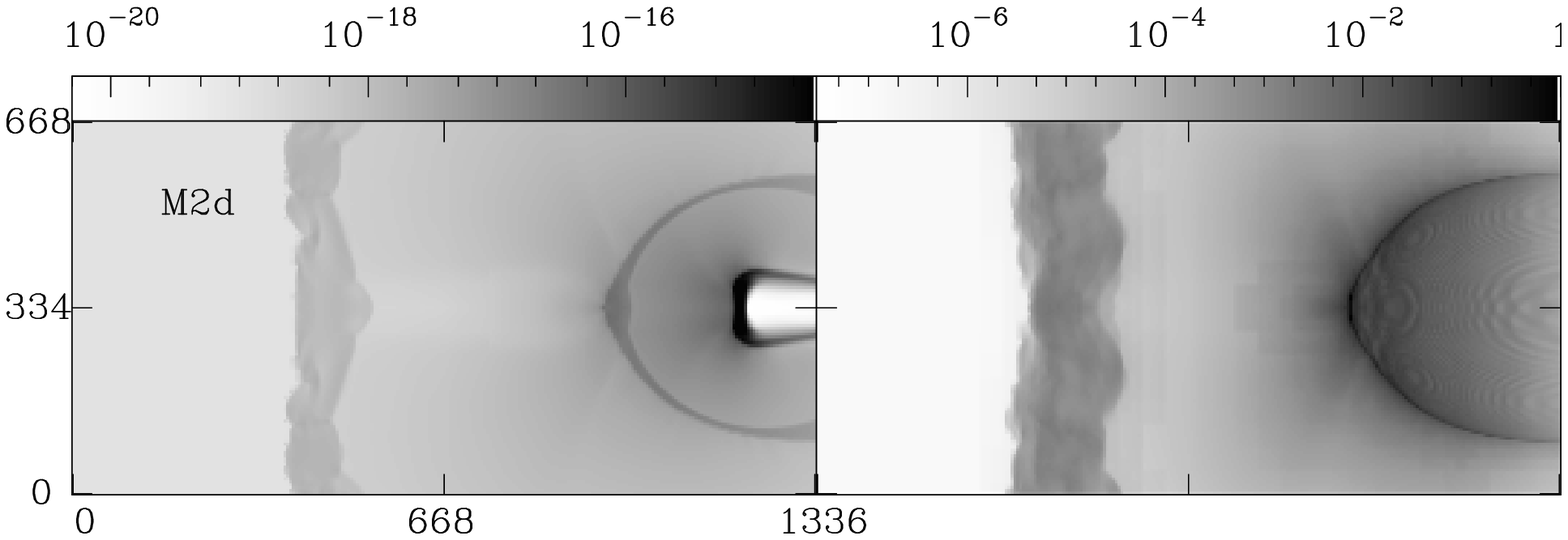} 
\caption{Density $xz$-midplane stratification (left) and H$\alpha$
map (right) for model M2d at $t=$ 200 yr. The temperature of the
PDR region in this model is $T_2=3\,000$ K. The scales (given by
the top bars) are in g cm$^{-3}$ for the density (left), and
normalized to one for the H$\alpha$ maps (right). The coordinate
axes are in AU.}
\label{f12} 
\end{figure*}

In Fig. \ref{f12}, we show the $xz$-midplane density stratifications
(left) and H$\alpha$ maps (right) for model M2d at $t=$ 200 yr.
This model has the same parameters as the M2b model ($\theta=0^{\circ}$,
$d=$ 0.1 pc; see Table \ref{tab1}) but a different PDR temperature
$T_2=3\,000$ K (see equation \ref{temp}).  The ionization front
radius is larger, by a factor of $\simeq 3$, for the model M2d when
compared with those with $T_2=1\,000$ K. The mass loss rate is
$\approx 5\times 10^{-7}$ M$_\odot$ yr$^{-1}$ for M2d (about $3.5
\dot{M}_{M2}$). We have run equivalent simulations for models M3b
and M4b, that is, a model with $\theta=45^{\circ}$ and $d=0.1$ pc,
but with $T_2=3\,000$ K, and a model with $\theta=75^{\circ}$ and
$d=0.1$ pc, but with $T_2=3\,000$ K. For these models (not shown
here), we found the same trend of larger mass loss rates and
ionization front standoff distances for larger values of $T_2$.

\section{Conclusions} \label{conclusions}

This work describes the first fully three-dimensional numerical
gasdynamic simulations of the photoevaporation of {\it neutral}
disks, subject to an interaction with the FUV/EUV photon flux and
the wind from an external O star.

With this initial setup, we study the time-evolution of the
photoevaporated flow. We have calculated models in the EUV dominated
regime, in which one obtains an ionization front (IF) close to the
surface of the disk, and a photodissociation region (PDR) not
resolved by our numerical scheme between the surface of the disk
and the IF, and in the FUV dominated regime, in which a PDR (well
resolved in our numerical simulations) is formed between the surface
of the disk and the IF. Both EUV and FUV dominated models develop
an interaction region between the photoevaporated wind and the wind
from the hot star. In some cases, this interaction region lies
within our computational domain.  For these cases, the interaction
region lies close to the position of stellar wind/photoevaporated
flow ram pressure balance.

We have simulated models of disks with different inclinations with
respect to the direction of the impinging EUV and FUV photon fluxes,
and at different distances from the ionizing photon sources ranging
from 0.02 pc to 0.2 pc.  The model at a distance of 0.02 pc from
the O-star is clearly EUV dominated, since the IF is just above the
disk surface.  Models at distances from 0.1 pc to 0.2 pc show
detached PDRs. These PDRs show structures, such as the presence of
a neutral supersonic flow near the disk and, farther out  but before
reaching the IF, a subsonic neutral flow, that are consistent with
the main features predicted by Johnstone et al.  (1998) for a
FUV-dominated flow. We note that these structures occur in our
simulations even for the inclined disk models.  The typical size
of the IF was found to be in agreement with analytical results.  In
particular, the limits predicted by \cite{john..98} for the ionization
front radius, $2.5r_d \lesssim r_{IF} \lesssim 4 r_d$, were recovered
in most of the models, with or without disk inclination.

We found that, for a given inclination of the disk, the mass loss
rate seems to be independent of the distance to the FUV/EUV photon
source (for the FUV-dominated flows). In our simulations this is
due to the constancy of the PDR temperature. For higher disk
inclinations, we obtain smaller mass loss rates, which is a direct
result of the smaller effective cross section presented by the disk
to the impinging FUV/EUV photon flux.  Taking the mean values of
the mass loss rate from our simulations, we obtain $\dot{M}_{M2} =
1.4\times10^{-7}$ M$_{\odot}$ yr$^{-1}$, $\dot{M}_{M3} = 6.2\times10^{-8}$
M$_{\odot}$ yr$^{-1}$$\dot{M}_{M4} = 4.8\times10^{-8}$ M$_{\odot}$
yr$^{-1}$, for models with $\theta=0^{\circ}$, $\theta=45^{\circ}$
and $\theta=75^{\circ}$, respectively. These values are surprisingly
similar, in spite of the very different inclination angles used in
the simulations.

Our models predict H$\alpha$ emission maps with moderate side-to-side
asymmetries, which directly result from the misalignment of the
axis of the accretion disk with respect to the direction to the
photoionizing source. Asymmetries are seen both in the emission
associated with the IF and in the emission of the photoevaporated
wind/expanding H\,II region interaction working surface.  It is
clear that asymmetries in these two structures are indeed observed
in many of the Orion proplyds \citep{bally..00}. A detailed attempt
to model a specific, asymmetrical proplyd with a 3D photoionizing
disk simulation will be presented in a future paper.

\begin{acknowledgements}
We thanks the anonymous referee for his/her comments and suggestions,
which contributes to improve the quality of the manuscript. MJV
would like to thanks FAPESB for partial financial support (PPP
project 7606/2006). AHC wish to thank the Brazilian agency CNPq for
partial financial support (projects 307036/2009-0 and 471254/2008-8),
and also PROPP-UESC (projects 635 and 802). AR acknowledges support
from the CONACyT grants 61547, 101356, and 101975. We thank A.
Esquivel for sharing with us a set of IDL routines for plotting the
YGUAZU-a output.
\end{acknowledgements}

\vfill
\eject
\null

\end{document}